# Fully 3D-Printed Organic Electrochemical Transistors

Matteo Massetti[1], Silan Zhang[1,2], Harikesh Padinare[1], Bernhard Burtscher[1], Chiara Diacci[1], Daniel T. Simon[1], Xianjie Liu[1], Mats Fahlman[1,2], Deyu Tu[1], Magnus Berggren[1,2], Simone Fabiano[1,2]*

[1]Laboratory of Organic Electronics, Department of Science and Technology, Linköping University, SE-601 74 Norrköping, Sweden. E-mail: simone.fabiano@liu.se

[2]Wallenberg Wood Science Center, Linköping University, SE-601 74 Norrköping, Sweden.

**Abstract**

Organic electrochemical transistors (OECTs) are currently being investigated for various applications, ranging from sensors to logics and neuromorphic hardware. The fabrication process must be compatible with flexible and scalable digital techniques to address this wide spectrum of applications. Here, we report a direct-write additive process to fabricate fully 3D printed OECTs. We developed 3D printable conducting, semiconducting, insulating, and electrolyte inks to achieve this. The 3D-printed OECTs, operating in the depletion mode, can be fabricated on thin and flexible substrates, yielding high mechanical and environmental stability. We also developed a 3D printable nanocellulose formulation for the OECT substrate, demonstrating one of the first examples of fully 3D printed electronic devices. Good dopamine biosensing capabilities (limit of detection down to 6 µM without metal gate electrodes) and long-term (~1 hour) synapses response underscore that the present OECT manufacturing strategy is suitable for diverse applications requiring rapid design change and digitally enabled direct-write techniques.

**Introduction**

Organic electrochemical transistors (OECTs) are an enabling technology for many (bio-)electronic applications. OECTs owe their success to several compelling features such as simple device structure, tunable electronic properties to match different applications, biocompatibility, and simple manufacturing[1-4]. OECTs consist of a patterned organic (semi-)conductor thin film connecting the source and drain electrodes. An electrolyte is in contact with the semiconductor channel and the gate electrode. Applying a voltage to the gate ($V_G$) induces ion injection from the electrolyte into the semiconductor channel, causing the latter to change its doping state and conductivity. This modifies the current flowing between the source and drain electrodes (drain current, $I_D$). Contrary to standard thin-film transistors (TFTs), the doping mechanism is volumetric (i.e., occurring throughout the bulk of the channel material), leading to large $I_D$ variations even for small changes in $V_G$. Therefore, OECTs have shown remarkably high transconductance ($g_m = \partial I_D / \partial V_G$), despite being typically slower than TFTs, as ions have to penetrate the entire channel layer.[5] Because of this, OECTs have been successfully applied in bioelectronics, exploiting their high transconductance for biosensing[6-10], and printable electronics for low-power devices[11,12]. In addition, recent studies have underlined the possibility of using OECTs to emulate the functioning of biological synapses[13-15] and neurons[16,17], offering a unique opportunity to develop neuromorphic hardware for beyond-von-Neumann computing applications.

While many recent works have demonstrated the possibility of fabricating OECTs using a variety of additive manufacturing techniques, such as screen printing[18,19], inkjet printing[20-22], aerosol-jet printing[23,24] and gel extrusion[6,25,26], photolithography remains by far the most common approach for fabricating OECTs and circuits thereof, with excellent yields, high resolution, and scalability[27-30]. There is, however, a growing need for rapid design changes and digitally enabled direct-write techniques to enable high throughput screening of new materials and device concepts. To this end, 3D printing offers unique advantages compared to commonly used screen printing, gravure printing, or inkjet printing, such as superior flexibility in terms of material selection, ink preparation, and pattern design/change[4]. 3D printing allows for the deposition of a broad spectrum of materials (from insulators to semiconductors and conductors) with a large range of viscosity[31], similar to screen printing techniques. However, compared to screen printing, 3D printing allows for much higher flexibility in terms of pattern modification, since no new screens need to be produced for any changes to the pattern. In addition, compared to inkjet printing, 3D printing enables the pattern of highly viscous plastic materials with a good resolution, allowing the integration of OECTs with, e.g., traditional microfluidics, thus opening new possibilities in the field of biosensing and bioelectronics in general[22,32,33]. Previous attempts to 3D print OECTs[6,25,26,34-36] have also used other manufacturing techniques because of difficulties in developing suitable metallic and semiconducting inks. In addition, no attempts have been made to create alternatives to liquid electrolytes, which is fundamental for integrating OECT into applications that might not be limited to simple demonstrators.

Here, we propose a simple direct-write additive manufacturing approach to fabricating OECTs, based on a wet extrusion-based 3D printing system (BIOX by Cellink). We developed 3D printable functional inks for each of the OECT's components (source/drain/gate electrodes, semiconductor channel, insulator, gate electrolyte, and substrate) and demonstrated the first fully 3D printed OECTs. PEDOT:PSS was chosen as the OECT channel material, a benchmark mixed ion-electron conducting polymer for OECTs[37,38]. The fully 3D printed OECTs show remarkable electrical and mechanical properties and can be used for biosensing and neuromorphic computing applications. As wet extrusion-based 3D printers can be used to pattern cells[39,40], tissues[41,42] and organs[43,44], we anticipate that the possibility to fully 3D printing OECTs will pave the way for easier interfacing of electronics with biology.

**Results and Discussion**

**OECTs Fabrication Process Overview**

**Figure 1** illustrates the OECT fabrication process and final device layout. The source, drain, and gate electrodes were 3D printed on a microscope glass slide coated with a thin (< 4 μm) parylene layer using a blend of graphene oxide (GO) and carbon nanotubes (CNT) (see Experimental section for details about the ink formulation). The printed GO/CNT electrodes were treated with 50 wt% potassium iodide (KI) in 1 M hydrochloric acid (HCl) overnight to yield reduced GO (rGO),[45] resulting in rGO/CNT electrodes with electrical conductivity as large as 600 S cm$^{-1}$ (see Figure S1, Supporting Information). The oxidation level of GO and rGO was measured by X-ray photoelectron spectroscopy (XPS) (see Figure S2a for the C1s spectra, Supporting Information). For GO, deconvolution of the C1s spectra shows three peaks corresponding to sp$^2$ (C=C, 284.8 eV), epoxy/hydroxyl (C–O, 286.2 eV), and carbonyl (C=O, 287.8 eV) carbon atoms[46]. In contrast, the C1s spectra of the rGO samples only show the presence of C=C and C–C (285.4 eV). The C/O atom ratio calculated from XPS survey spectra also increases from 1.55 for GO to 7.38 for rGO, indicative of an effective GO reduction[47]. Ultraviolet photoelectron spectroscopy (UPS) was used to evaluate the work function (WF) changes of GO, rGO, and rGO/CNT (Figure S2b, Supporting Information). The WF of GO was measured to be about 5.05 eV and decreased to 4.34 eV after reduction to rGO. As the WF of the pristine CNTs is about 4.8 eV,[48] a final WF of 4.54 eV for rGO/CNT is indicative of a homogeneous dispersion of rGO and CNTs. To further decrease the gate electrode's electrical resistance and increase its ion injection capability and switching speed, we 3D printed a high conductivity PEDOT:PSS formulation (4 wt%, ≈ 200 S cm$^{-1}$, Figure S1) on top of the rGO/CNT gate electrode. The high conductivity PEDOT:PSS formulation was prepared following a procedure reported previously[49].

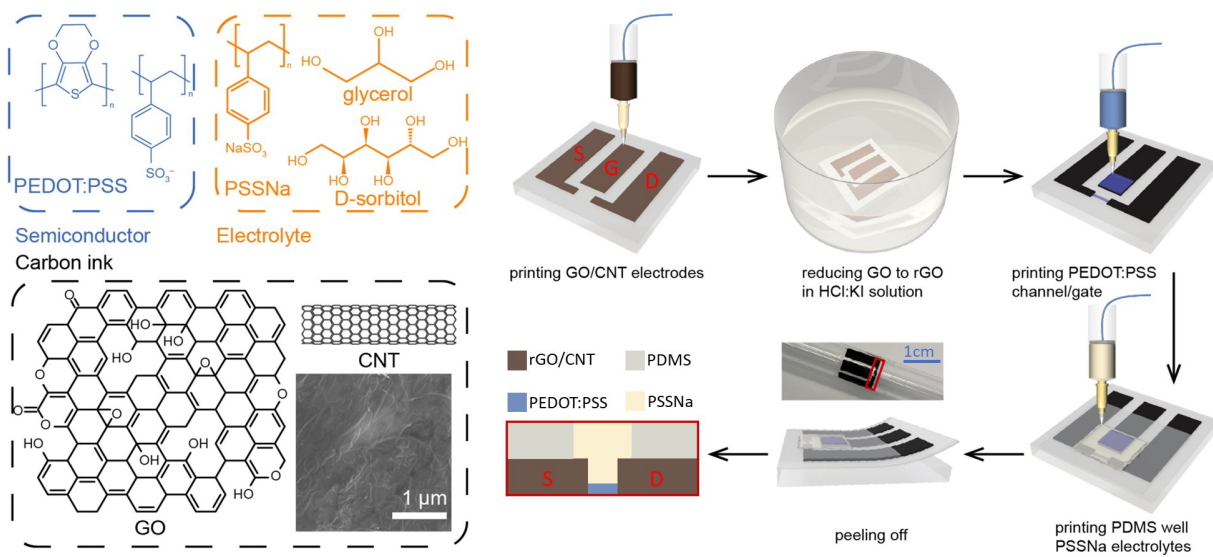

**Figure 1** Schematic of the OECT fabrication process and the materials employed.

Having defined the source, drain, and gate electrodes, we 3D printed the OECT channel using another PEDOT:PSS formulation having a lower concentration (i.e., 2.5 wt%). To ensure fast OECT response time and low operational voltage, we added an excess of D-Sorbitol (20 wt %) to the PEDOT:PSS formulation. D-Sorbitol is a well-known ion reservoir capable of maintaining mobile ions inside the semiconducting polymers and a suitable plasticizer for PEDOT:PSS, promoting its adhesion to parylene.[50,51] Additionally, we used divinyl-sulfone (DVS) to boost the PEDOT:PSS channel water stability, without sacrificing its electrical conductivity.[52,53] A PDMS insulating layer with an open area of about $0.5 \times 3$ mm$^2$ was then printed on top of the PEDOT:PSS channel (width $W = 240$ μm and length $L = 77$ μm), and the gate electrode ($2 \times 6$ mm$^2$) to define a well for the electrolyte. For the latter, we printed a poly (sodium 4-styrenesulfonate) (PSSNa) based hydrogel (50 wt%) containing 0.2 M aluminum chloride (AlCl$_3$) to increase its ionic conductivity through the addition of more ionic species (Figure S3, Supporting Information). The fully printed OECTs, possessing a footprint of $7 \times 7$ mm$^2$, are lightweight and show good mechanical flexibility (see picture in Figure 1).

**3D-Printed OECT Characterization**

The electrical characteristics of the fully 3D printed PEDOT:PSS-based OECTs are reported in **Figure 2**a (W × L = 77 μm × 240 μm). The transfer curves, recorded at drain voltage $V_D$ = -1 V, are typical for a depletion mode PEDOT:PSS-based OECT, with an ON/OFF current ratio exceeding 2800, at gate voltage $V_G$ < 1 V and a geometry-normalized transconductance ($g_m$) as high as 34 S cm$^{-1}$. The latter is comparable to typical values of (inkjet, screen, and aerosol-jet) printed PEDOT:PSS-based OECTs.[18,24] The transfer curves also show low hysteresis (< 20 mV), indicative of easily reversible dedoping/doping processes. The output characteristics are also consistent with typical PEDOT:PSS-based OECTs (Figure S4a, Supporting Information). The transient response of our 3D-printed PEDOT:PSS-based OECTs, measured using gate voltage pulses of $V_G$ = 0.8V, is reported in Figure S4b. The switching off ($\tau_{OFF}$) and switching on ($\tau_{ON}$) times were extracted by fitting the $I_D$ temporal response with a single exponential decrease function (Figure S4c, Supporting Information), yielding a $\tau_{OFF} \approx$ 27 ms and a $\tau_{ON} \approx$ 100 ms. These results are remarkable considering they refer to fully 3D-printed OECTs and are comparable to screen-printed devices (see Table S1 in the Supporting Information).[18,19,54]

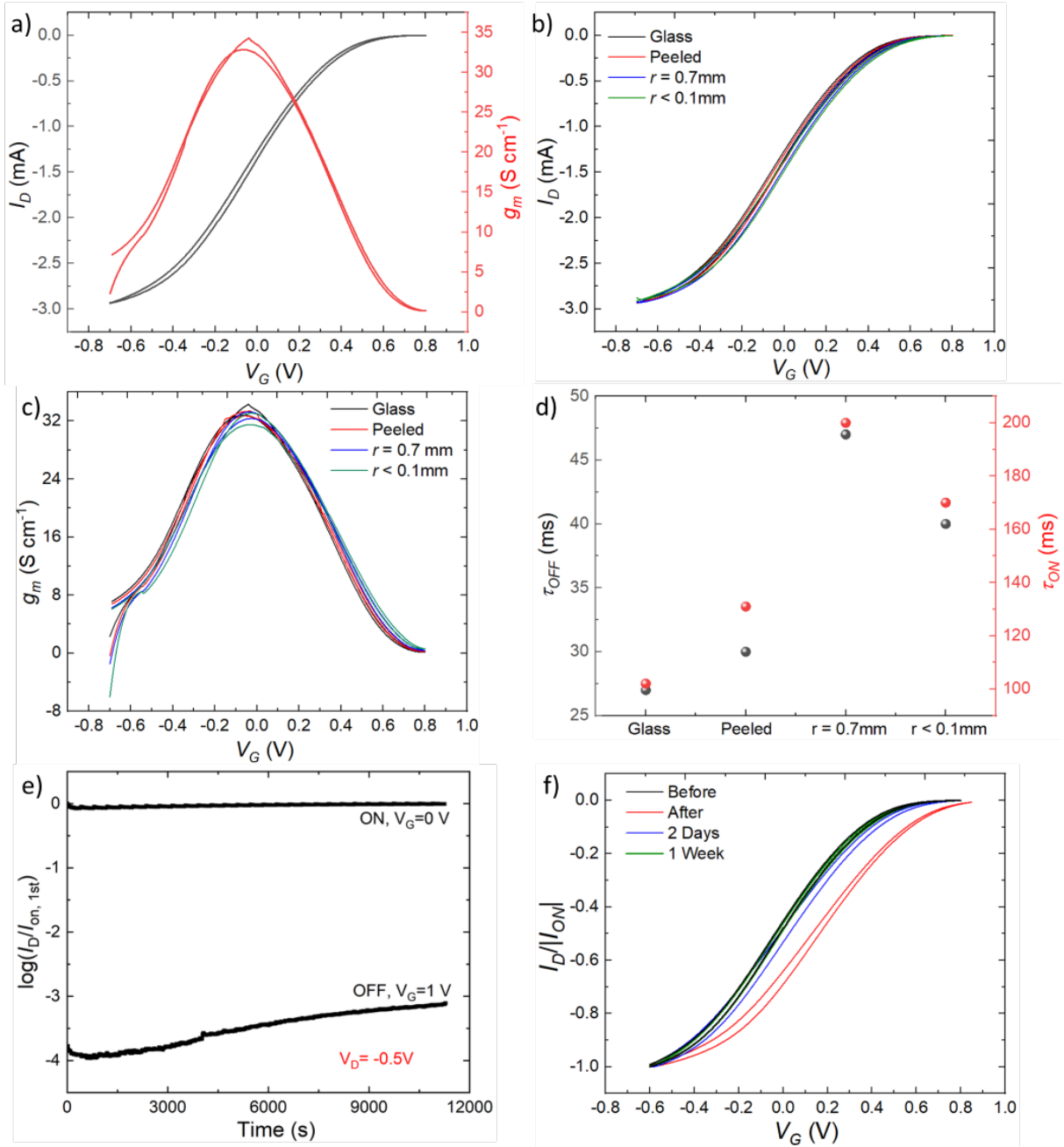

**Figure 2.** a) Transfer curve and normalized transconductance ($g_m$) of the 3D printed OECT. OECT transfer curves b) and transconductances c) under different mechanical stress. D) OECT response time under different mechanical stress. E) OECT electrical stability under $V_G$ cycling (0-1 V) for over 3 h. f) Comparison of the normalized transfer curves before and after the cycling stability and their recovery after resting at 6 °C and high humidity.

To investigate if the channel or the electrolyte material caused the asymmetry in ON-to-OFF and OFF-to-ON switching speeds, we performed the same measurements using 0.1 M NaCl liquid electrolyte. The results are reported in Figure S5 (Supporting Information) and show $\tau_{OFF}$

= 7.5 ms and $\tau_{ON}$ = 7.8 ms when aqueous NaCl is used as the electrolyte. The similar $\tau_{OFF}$ and $\tau_{ON}$ values suggest that ion trapping inside the PEDOT:PSS channel occurs when PSSNa+AlCl$_3$ based hydrogel is used as the gel electrolyte, likely due to the trapping of large ionic species inside the semiconducting channel material. Note that a similar asymmetry in the switching speeds has already been observed in OECTs and linked to steric hindrance of the counterions[50,55,56] or the electrolyte ions dimensions.[57,58]

As the devices are fabricated on top of a thin (< 4 μm) parylene layer, they can be easily peeled off from the glass slide. These free-standing devices can be wrapped around a glass pipette (radius $r$ = 0.7 mm) and even folded in half, perpendicularly to the channel (radius $r$ < 0.1 mm), without any appreciable changes in the electrical performance, as also visible from the transfer and transconductance curves reported in Figure 2b-c. We observed a < 2× decrease in the switching speed of the bent device compared to the unfolded devices. However, since the ON-state current is the same for all cases, we speculate that the changes in device time response are due to partial delamination of the gel electrolyte from the PEDOT:PSS channel. Despite the slightly slower switching speed, our devices show remarkable mechanical stability even under high stresses.

Finally, we tested the electrical performance stability of our 3D-printed PEDOT:PSS OECTs under 3h continuous operation in ambient at pulsed $V_G$ = 1 V (Figure 2e) and observed a 10× increase of the OFF current (from 0.1 μA to about 1 μA after 3 h) as well as a small 20% increase in the ON current (from 1 mA to 1.2 mA after 3 h). To better understand the origin of the small rise in $I_D$, we measured the transfer curves before and after the cycling test (Figure 2f, unnormalized curves in Figure S6 Supporting Information). When the transfer curves are normalized to the ON current, a less effective gate modulation can be observed, with the OECT not completely turning OFF at $V_G$ = 0.8 V (ON/OFF ratio decreases from 2400 before stress to 140 after the cycling stress). However, the electrical characteristics are partially recovered after storing the devices in the fridge for 2 days (ON/OFF ratio > 1000) and are completely recovered after one week of storage inside the refrigerator. Therefore, we speculate that some of the water inside the gel electrolyte evaporated due to heating during continuous operation, leading to a decrease in ion mobility inside the electrolyte. The gel could absorb humidity from the environment by storing the device inside the fridge, thus re-establishing an environment favorable to ion motion.

**Beyond Traditional Printing Techniques**

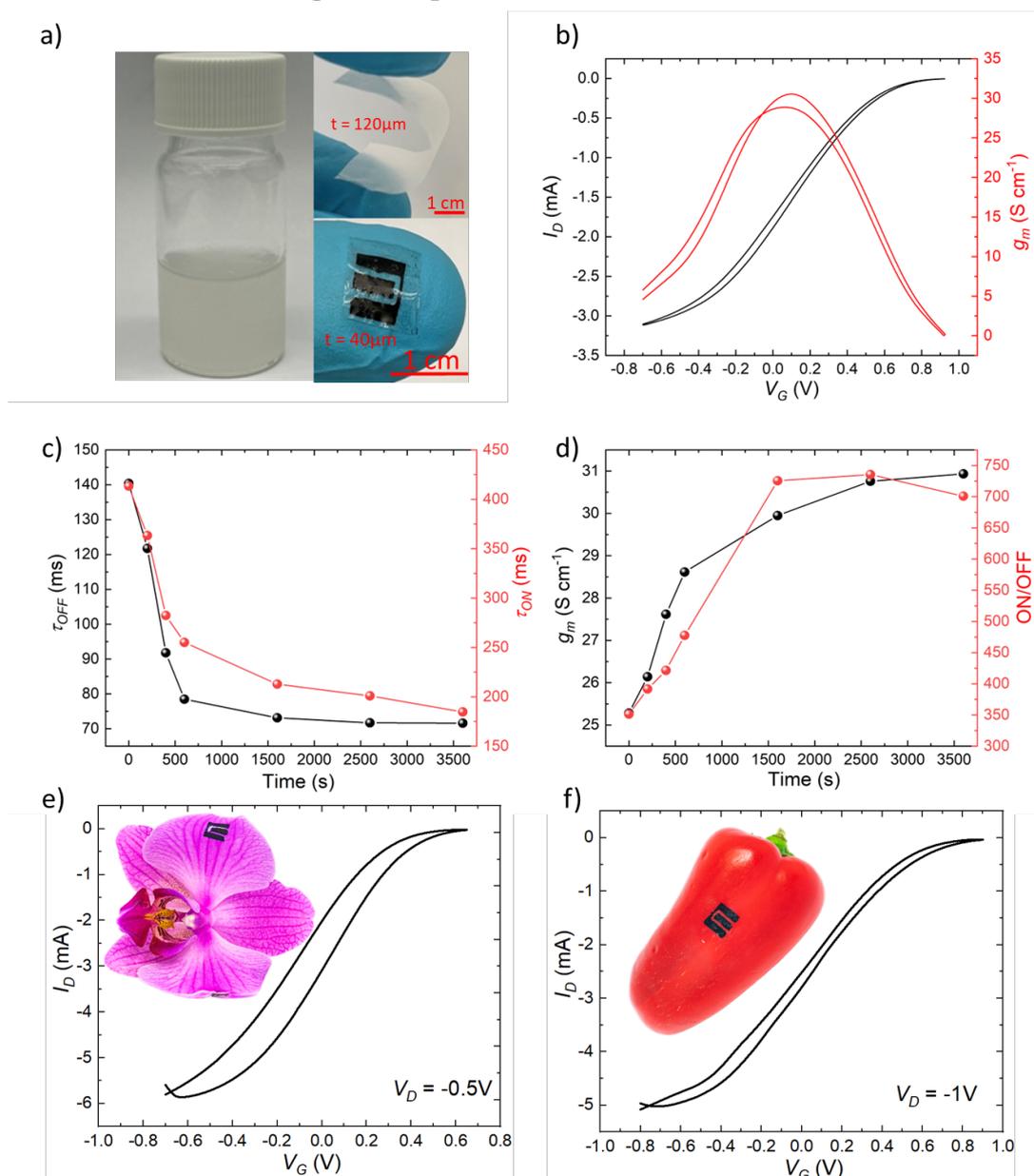

**Figure 3.** a) Photos of the cellulose ink (left), printed film (top right) and OECT (bottom right). b) OECT on cellulose transfer curve and normalized transconductance ($g_m$). c) Evolution of the time response ($\tau_{OFF}$ and $\tau_{ON}$) after biasing the gate at -0.7 V for different times (up to 1 h in total). d) Evolution of the $g_m$ and ON/OFF ratio after biasing the gate at -0.7 V for different

times (up to 1 h). Transfer curves of OECTs directly printed on e) an orchid flower and f) a bell pepper.

While in the previous section, we explored the possibility of printing all the elements of an OECT using 3D printing, here we highlight the advantages of this fabrication process compared to more traditional printing techniques. Other ways of patterning devices would allow for higher throughput (e.g., screen printing or gravure printing) or higher resolution (e.g., inkjet/aerosol jet printing and standard photolithography/spin coating). However, high throughput deposition/patterning techniques are severely limited to flat, flexible, or stretchable substrates with minimal possibility of dealing with unconventional (irregular, curved, or soft) surfaces[59]. On the other hand, techniques such as inkjet or spin-coating are more limited in terms of processable materials, only allowing the patterning of low-viscosity functional inks. While a few reports have shown the potential of printing devices on very thin substrates, which are then laminated or wrapped around more complex objects, after processing,[60-62] direct deposition/patterning on complex structures using these techniques is typically troublesome. On the contrary, exploiting the high viscosity of the inks, it is possible to 3D print structures on unconventional objects with irregular surfaces. Thus, to demonstrate the full potential of 3D printing, we targeted two different objectives: 1) the printing of a complete device, including a flexible and self-standing substrate, and 2) the patterning on irregular surfaces.

Regarding the first point, while using parylene allows us to fabricate flexible free-standing devices, it clashes with the idea of devices manufactured using low-cost printing techniques. Indeed, deposition of parylene requires vacuum systems and a high temperature of > 600 °C, with most of the material wasted inside the deposition chamber. Therefore, we exploited the possibility of our 3D printing system to print highly viscous formulations to also directly print the OECT substrate. With the aim to use only low-cost materials/processes, we developed a nanofibrillated cellulose (NFC)/polyvinyl alcohol (PVA) (1:1 ratio) hydrogel that can be easily processed at low temperatures (< 100°C). To obtain a water-insoluble substrate, we added glutaraldehyde to our formulation, which is known to crosslink PVA after acid treatment.[63,64] In addition, the crosslinking of the substrate is mediated by the presence of an acidic environment, which allows us to use the same solution used to reduce the GO electrodes to process both substrate and electrical connections at the same time. First, we 3D printed the NFC/PVA substrate and dried it in an oven at 80 °C for 1 h. The resulting 60-µm-thick film was then used as the substrate upon which we 3D printed the GO/CNT electrodes. Then, we immersed the NFC/PVA film and GO/CNT electrodes overnight in the reducing solution, as described in Section 2.1. After removing the iodine contaminants in the rGO/CNT electrodes, by immersing the substrate and electrical connections in ethanol, the devices were dried inside an oven at 80 °C for 2 h. The subsequent channel and electrolyte layers were deposited following the process described in Section 2.1.

**Figure 3a-d** shows the electrical characteristics of the OECTs 3D printed on NFC/PVA substrates. The ON current matches the $I_D$ values recorded for the OECTs fabricated on parylene, indicative of an excellent reproducibility of the printing process. The major differences, however, concern the OFF-state current and switching characteristics. We observed a significantly reduced ON/OFF ratio of 700 against 2400 of the OECTs printed on parylene and a slower temporal response with $\tau_{OFF} \approx 70$ ms and $\tau_{ON} \approx 170$ ms (Figure 3b). We attributed the slightly lower performance to the use of anionic NFC as the substrate material, which is known to be a good ionic conductor.[65,66] We then speculate that part of the ions is injected from the gel electrolyte into the underlying cellulose substrate. This seems to be confirmed by the fact that applying a negative gate bias (-0.7 V) positively affects the ON/OFF ratio and time response (Figure 3c-d). This bottleneck could be solved in the future using cationic cellulose-based materials, thus preventing the injection of cations into the substrate.

As for the possibility of printing devices onto unconventional substrates, we 3D printed the OECTs on soft, curved, and irregular surfaces such as those of red bell peppers and orchid flowers. Given the fragile nature of these substrates, we modified the printing process by replacing the rGO/CNT ink with the high conductivity PEDOT:PSS formulation used on the gate of the OECTs described above. This is because reducing the rGO/CNT layers requires acid treatment and a heating step that would otherwise compromise the integrity of the plant-based substrate. **Figures 3e-f** show the electrical characteristics of the OECTs 3D printed on orchid petals and bell pepper. Because of the rough and complex substrate surface, we used a larger channel area and channel thickness for this type of device. For this reason, the $I_{D,ON}$ is higher for both devices than those printed on parylene and paper, with an ON/OFF ratio of around 300 (orchid) and 160 (bell pepper). We ascribed the lower ON/OFF ratio to two major factors: 1) a thicker channel layer and 2) the soft/ion permeable nature of the plant-based surface, similar to the NFC substrate. In addition, we estimated device response times in the range 0.5-1 s (Figure S7, Supporting Information), which are expected considering the thicker channel layers and the use of electrochemically active PEDOT:PSS source/drain electrodes[67,68]. Despite the somewhat lower performance of these unoptimized devices compared to OECTs 3D printed on standard parylene substrates, pattering working devices on complex surfaces—impossible with traditional printing techniques and photolithography—have significant implications for the field of biosensing and edible electronics where modulation speed is not always a stringent requirement. Future optimization should focus on developing milder curing/post-processing conditions for reducing the rGO/CNT electrodes.

**Applications of 3D-Printed OECTs**

Having characterized the electrical properties of the 3D printed OECTs, we tested these devices for two representative applications: biosensing and neuromorphic.

*BIOSENSING: Dopamine Detection:* First, we evaluated the capability of our 3D-printed OECTs to detect biomolecules of interest in neuromedicine. Specifically, we tested the detection of dopamine (DA), a neurotransmitter regulating the correct functioning of several organs and whose loss in some areas of the brain is linked to, e.g., Parkinson's disease. Another motivation for choosing DA detection is that DA is a benchmark molecule for testing OECT biosensing due to its easy detection mechanism[69-72]. The reduction of dopamine at the gate electrode modulates the effective $V_G$ due to an offset voltage strictly connected to the analyte concentration, which modulates the $I_D$[7]. We modified our OECT structure for DA detection by removing the PEDOT:PSS layer at the gate electrode. The reason is to shift the transconductance peak at higher values of $V_G$, to have dopamine reduction at the gate, which generally happens at $V_G > 0.4$ V[73-75].

The 3D printed OECT response to changes in DA concentration and relative calibration curve are reported in **Figure 4**. To minimize device differences and measurement noise, we used the normalized response (NR) against the fit curve (Figure 4a):

$$NR = \Delta I/I = -(I_{DA} - I_{Fit})/I_{Fit}$$

where $I_{Fit}$ is the fitted $I_D$ of the blank buffer and $I_{DA}$ is the drain current in the presence of dopamine. To obtain the dose curve, the dopamine concentration was changed from 1 µM to 10 mM every 100 s and the value before switching to a higher concentration was taken for the dose curve.

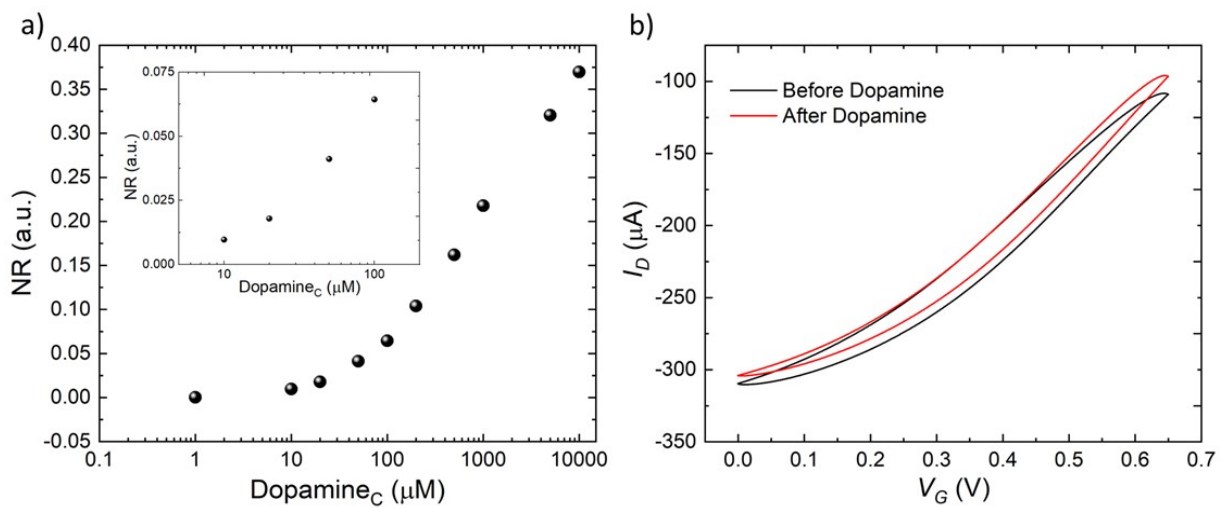

**Figure 4**. a) Dose curve for dopamine detection (inset is the focus on the 10-100 µM region). b) Transfer curves before and after the dopamine detection experiment.

Both samples stably followed the same trend when the PBS buffer solution was changed multiple times. With increasing the DA concentration, a rise in the normalized response (decrease in $I_D$) is observed due to a faradaic current at the gate electrode. The devices showed a fast response to changes in DA concentration and exhibited a near-step-like behavior. While the devices showed a linear behavior at low DA concentrations (up to about 100 µM), they started saturating at higher DA concentrations (above 1 mM). For these concentrations, a maximum NR of about 0.4 is achieved, leaving ample room for improvement. For example, a common strategy to be employed could be the reduction of the gate dimension, which was shown to increase sensitivity in sensing peroxide[76]. This is due to a more significant potential drop over the gate/electrolyte interface compared to the larger gate dimension. The transfer curves before and after dopamine sensing exhibit similar characteristics (Figure 4b). The device showed a slight $I_D$ shift in transfer sweeps (at $V_G = 0$ V) before and after DA sensing. This could be attributed to device bias stress during measurement or irreversible interaction of DA with the gate electrode (e.g., through the formation of polydopamine[73-75]). The latter could change the transfer curve by altering the capacitance at the gate/electrolyte interface.

To obtain the limit of detection (LOD), i.e., the lowest DA concentration our devices can detect, we used a linear regression model[77]. As LOD is equal to $3.3 \times S_a/b$, with $S_a$ the response standard deviation and $b$ the curve's slope in the linear regime of the dose curve, we extrapolated a value of 6 µM. This value is remarkable considering the absence of a metal gate (e.g., Pt) typically used to detect lower dopamine concentrations and is well within physiological values[8,35,69,70]. To further improve the LOD, future investigation should focus on optimizing the gate functionalization with, e.g., Pt nanoparticles and improving the gate/channel geometry, thus increasing the S/N ratio and device sensitivity.

*NEUROMORPHIC: PEDOT:PSS Channel Long-Term Depression:* Synapse is the connection between two neurons and enables information exchange through chemical and electrical signals.

Learning in the human brain is caused by the permanent/temporary strengthening and weakening of synaptic connections. The strength of these connections is referred to as the synaptic weight and is modulated by processes called potentiation and depression. Long-term changes in the synaptic weight (minutes to hours or longer), called long-term potentiation/depression (LTP/LTD), are the basis for experience-dependent alteration of the neural network. In contrast, short-term changes in synaptic weights (ms to s), known as short-term potentiation/depression (STP/STD), assist in decoding temporal information[78].

OECTs have been already employed to emulate synaptic functionality[14,79-81] and even to simulate neural behavior,[16,17]. Still, to our knowledge, no previous report of fully 3D-printed devices showing neuromorphic capabilities, with state retention longer than 1 hour and employing a gel-based electrolyte. In particular, using a non-liquid electrolyte could push the integration of neuromorphic devices with other electronic systems leading them closer to real-life applications.

Analogous to the biological synapses, in the 3D printed OECT-based artificial synapses, the voltage input to the gate represents the presynaptic input, and the current across drain and source is the postsynaptic output referred to as the excitatory postsynaptic current (EPSC). The artificial synapse also operates in short-term and long-term plasticity modes similar to the biological counterpart. Ionic buildup in the PEDOT:PSS channel during gate-induced doping/dedoping causes a short-term increase/decrease in channel conductance over several seconds (**Figure 5**a). Short-term synaptic facilitation (ms to s) in biological synapses enabled by the accumulation of excess $Ca^{2+}$ ions and neurotransmitters on facing recurrent action potentials from the neuron is analogous to this process. The EPSC can be modulated by changing the gate bias from -0.4 V to -0.8 V, resulting in an enhanced excitatory effect.

Paired pulse facilitation (PPF) is a type of short-term plasticity where the EPSC increases progressively with repeated action potentials. PPF is involved in decoding temporal information in biological systems. PPF index is the ratio of the amplitudes of the second and the initial

EPSC on the application of consecutive presynaptic inputs. PPF indices of the artificial synapse exhibited an exponential decay with increasing pulse intervals resembling the signaling characteristics of biological synapses (Figure 5b-c). It can be fitted with the exponential decay equation:

$$y = A_1 \times \exp\left(\frac{-x}{t_1}\right) + A_2 \times \exp\left(\frac{-x}{t_2}\right) + c$$

where x represents pulse interval duration, c = 100.38 represents resting magnitude, $A_1$ = 12.31 and $A_2$ = 9.92 are the facilitation constants, and $t_1$ = 50 ms and $t_2$ = 134.5 ms are the characteristic time constants. These characteristic time constants closely resemble the values occurring in biological synapses (40 ms and 300 ms)[81].

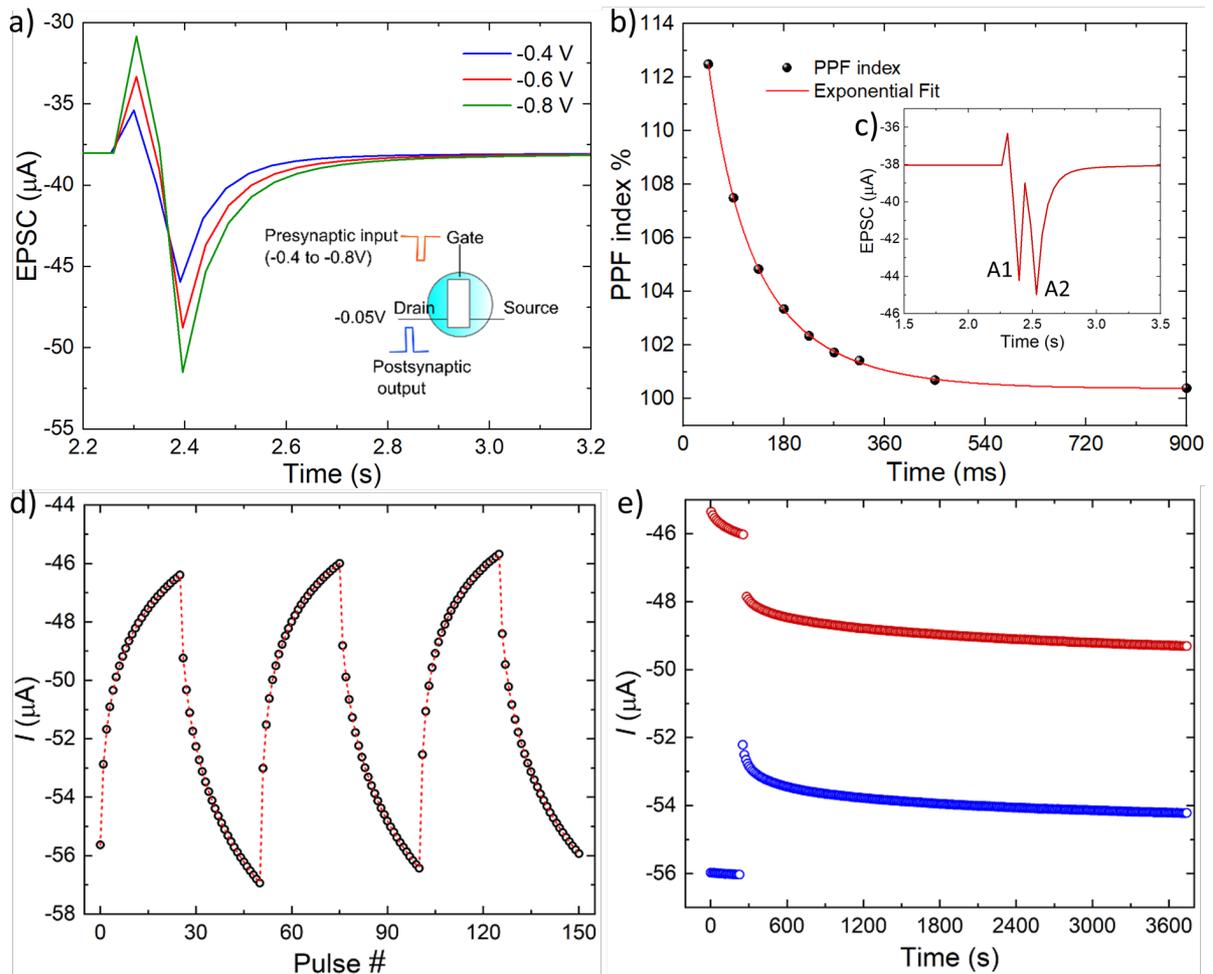

**Figure 5.** The Excitatory postsynaptic current measured at the drain (-0.05 V) on application of various gate pulse voltages of 90 ms duration. (b) The paired-pulse facilitation indices of the synapse operating in the short-term doping-dedoping mode. PPF index is given by (A2/A1) × 100 (c) The increase in EPSC on application of consecutive gate pulses separated by Δt. (d) Long-term depression and potentiation in the synapse exhibiting 25 distinct levels. Depression

is induced by 1.1V pulses of 2.5 s duration and potentiation by -0.6 V pulses of 4s duration. The time delay between two consecutive write pulses is 20 s. The conductance state is read using a drain voltage pulse of -0.05 V, 10 s after writing. (e) Retention of two states A and B over 1 hour.

Increasing the gate pulse width and amplitude resulted in long-term changes in the conductivity of the OECT channel. This is similar to the long-term plasticity in biological synapses, where the insertion of new receptors in synapses leads to permanent increases in synaptic strength. LTP is induced by -0.6 V pulses of 4s duration and LTD by application of 1.1 V pulses for 2.5 s (Figure 5d). 25 distinct conductance states are achieved in the synapse with retention of around 1 hour (Figure 5e). This is comparable to state-of-the-art OECT-based synapses[79,80] and does not require a floating gate setup to enable retention. The reason for the prolonged retention in the system is currently elusive and is under investigation. However, we hypothesize that it might be due to ion trapping in the 3D printed channel.[79]

In conclusion, we showed that a digital approach to additive manufacturing, extrusion-based wet 3D printing, can be used to fabricate OECTs rapidly. The fully 3D printed OECTs showed remarkable electrical characteristics, good biosensing capabilities (without using standard metal gates for detection), and neuromimicking paired-pulse depression behavior. Due to the mechanical flexibility of our 3D printed OECTs, we demonstrate the possibility of integrating biosensing and neuromorphic applications into personalized, low-cost devices fabricated by additive manufacturing.

Wet 3D additive manufacturing systems are still new to organic electronics since they are primarily employed in microfluidics and for printing biological structures[40,82-85]. 3D printing combines the advantage of patterning electronic devices on complex structures with flexibility in the device pattern design and low material waste (typical of drop-on-demand techniques). These features place 3D printing in a privileged position compared to traditional manufacturing techniques like screen or gravure printing, suffering from limited flexibility in pattern adjustment and higher material waste—a critical issue concerning the high material costs. The same applies to standard photolithographic patterning. While photolithography is a benchmark technique for the semiconductor industry, it is not suited for low costs processing. In addition, photolithography is usually coupled to spin coating for deposition of the organic semiconductors, another technique where >95 % of the semiconductor is wasted. The work presented here demonstrates the possibility of optimizing materials to fabricate flexible and self-standing OECTs demonstrating good electrical performances, good detection of biological signals, and neuromorphic behavior. This works thus shows a clear path for translation to more complex structures and devices that can then progress to applications in bioelectronics[1,86].

**Methods**

*GO synthesis*: All chemicals, including graphite flakes (~325 mesh), sulfuric acid ($H_2SO_4$, 98%), potassium permanganate ($KMnO_4$), hydrochloric acid (HCl), and hydrogen peroxide ($H_2O_2$), were obtained from Sigma-Aldrich and used as received. GO samples were prepared as reported in Ref. [46]. In brief, 3 g graphite flakes were dispersed in 90 mL 98% $H_2SO_4$ in a 250 mL flask and kept stirring at room temperature for 30 min. Then 9 g $KMnO_4$ was slowly added to the flask under vigorous stirring. The oxidation process was performed at room temperature for 3 h. After that, the reaction was determined by pouring the reaction system into 500 mL deionized water with 10 mL $H_2O_2$ to reduce Mn (VII) species. The suspension was stirred for 10 min and allowed to stand overnight. Then the supernatant was decanted and washed with 10 vol% HCl solution three times by centrifuge (10,000 rpm for 5 min) to remove sulfate ion. Three times washing with deionized water was followed by centrifuge (10000 rpm for 50 min) to remove any acid residue.

*GO/CNT ink preparation*: The GO/CNT ink was prepared by dispersing 150 mg GO and 75 mg CNT (~50 μm purchased from Jiangsu XFNANO Materials Tech Co., Ltd) into 10 ml of 50 vol% dimethyl sulfoxide (DMSO)/$H_2O$ and ultrasonicated for 6 hours.

*Conductive PEDOT:PSS ink preparation*: 100 ml of PEDOT:PSS commercial formulation (PH1000, Heraeus) was frozen in liquid nitrogen and then inserted in our freeze-dryer system (BenchTop Pro, SP-Scientific), where it was left for 72 hours to eliminate all the solvent inside, following the procedure described in Ref. [49]. Then the dry PEDOT:PSS was redispersed in 4wt% in a DMSO/$H_2O$ (vol% 5/95). To increase the plasticity of the formulation, 1wt% of D-Sorbitol and 4w% of Triton X were added. The components were mixed first using a homogenizer (T 10 basic ULTRA-TURRAX®, IKA) for 5 minutes and then magnetically stirred overnight at 120°C on a hotplate. Afterward, (3-glycidyloxypropyl) trimethoxysilane (GOPS, 0.1 wt%) was added, and the solution was then placed in a vacuum chamber for 30 minutes to remove all the air bubbles.

*OECT Channel PEDOT:PSS ink preparation*: 100 ml of PEDOT:PSS commercial formulation (PH1000, Heraeus) was frozen in liquid nitrogen and then inserted in our freeze-dryer system (BenchTop Pro, SP-Scientific), where it was left for 72 hours to eliminate all the solvent inside, following the procedure described by [ref]. Then the dry PEDOT:PSS was redispersed in 2.5 wt% in a DMSO/$H_2O$ (vol% 5/95). To increase the plasticity of the formulation, 2 wt% of Triton X was subsequently added. In addition, 20 wt% of D-Sorbitol was added to increase plasticity and promote ion diffusion inside the PEDOT:PSS film. The components were mixed first using a homogenizer (T 10 basic ULTRA-TURRAX®, IKA) for 5 minutes and then magnetically stirred overnight at 120 °C on a hotplate. Afterward, glycerol (1 wt%) and divinyl sulfone (DVS, 3 vol%) were added to the solution to have $PSS^-$ cross-linking at room temperature without losing electrical conductivity.[52,53] Finally, the solution was placed in a vacuum chamber for 30 minutes to remove all the air bubbles.

*Cellulose ink preparation*: The cellulose ink was fabricated as reported in Ref. [87]. In brief, 51.35 g carboxymethylated nanofibrillated cellulose (1 wt% solid content, ordered from Research Institutes of Sweden(RISE)), 11 g 2 mg mL$^{-1}$ polyvinyl alcohol (PVA, Mowiol® 18-88, Sigma-Aldrich) solution and 44 mg 25 wt% glutaraldehyde solution (Sigma-Aldrich) were homogenized together using a laboratory mixer (T 10 basic ULTRA-TURRAX®, IKA) for 10 minutes and then followed by degassing in a vacuum desiccator overnight to remove excessive bubbles in the viscous mixture.

*PSSNa gel electrolyte ink preparation*: 2.5 g of PSSNa (Sigma, Mw 2000000) and 0.5 g of D-Sorbitol were dispersed in 5 ml of 0.2 M of aluminum chloride ($AlCl_3$) and 10 ml of glycerol. The components were magnetically stirred overnight at 150°C on a hotplate and then at room temperature for an entire day to remove all the air bubbles inside the gel.

*PDMS ink preparation*: Two batches of different PDMS formulations, 10 g each of Sylgard 184 (Sy 184) and SE 1700 (Dow Corning, Auburn, MI), were separately prepared by mixing with their respective curing agent in a 10:1 ratio. Each blend was then mixed separately to ensure the proper dispersion of the curing agent (ARE-250 CE, Thinkymixer). Then the Sy 184 and SE 1700 were combined with a 6:4 ratio and centrifuged to ensure a homogeneous dispersion. Afterward, isopropyl alcohol (IPA) was added in a 1:2 ratio as thinning agent; the final formulation was then centrifuged once again to obtain a homogeneous 3D ink.

*OECT 3D-Printing.* The devices were fabricated using a commercial extrusion-based 3D-Printing system (BIOX by Cellink). The patterns employed to fabricate the OECTs have been drawn using Blender Software and saved as .stl files. These files were then sliced using the inbuilt BIOX software. The printing parameters for each material are here reported:

- GO/CNT. Nozzle: 30G. Pressure: 70Pa. Speed: 10 mm/s. Resolution between 120 to 150 μm
- PEDOT:PSS for the channel. Nozzle: 34G. Pressure: 80Pa. Speed: 5mm/s. Resolution between 100 to 120 μm
- PEDOT:PSS for the gate. Nozzle: 30G. Pressure: 200Pa. Speed: 15mm/s. Resolution 150 μm
- PDMS. Nozzle: 30G. Pressure: 200Pa. Speed: 15mm/s. Resolution 150 μm
- PSSNa electrolyte. Nozzle: 32G. Pressure: 500Pa. Speed: 10mm/s. Resolution 120 μm
- Cellulose. Nozzle: 27G. Pressure: 70Pa. Speed: 25mm/s. Resolution 200 μm

These parameters result from an optimization process to ensure a good filament extrusion and pattern infill with the best resolution. Higher pressure values caused material over deposition (thus, lower resolution), while lower pressure led to a non-completely filled pattern due to nonoptimal filament extrusion. Regarding the printing speed, higher values led to a nonoptimal filament deposition yielding an incomplete pattern, while lower values contributed to material spreading and thus reduced resolution.

*OECT 3D Printing of unconventional surfaces:* All-PEDOT:PSS-based OECTs were directly printed on orchid and bell pepper surfaces. The S, D, and G electrodes were fabricated using the conducting PEDOT:PSS formulation used for the gate electrode. The channel areas were 150 μm x 800 μm (orchid) and 300 μm x 800 μm. The PEDOT:PSS channel formulation was then printed inside the channel area. The OECTs were inserted in a vacuum desiccator for 2 hours to dry and then removed and left in air for 2 hours to remove any solvent from the printed formulations entirely. Before measuring the OECT on the orchid, the petals were placed for 15 minutes on a hotplate at 100°C to remove any liquid inside the plant that would cause large parasitic currents. Finally, the PSSNa electrolyte was 3D-printed on the devices to obtain the complete OECT structure.

*OECT and electrical conductivity measurements*: Electrical conductivity measurements of the semiconducting/conducting materials and OECT testing were performed using a 4200-SCS Semiconductor Characterization System. The obtained data were then analyzed and fitted using Python software.

*Ionic conductivity measurements*: The ionic conductivity was measured by electrochemical impedance spectra (Bio-logic Instrument) as reported previously[88]. The NaCl or PSSNa gel was modified with an area of 0.071 cm$^2$ and 0.35-cm-thick PDMS well and then sandwiched between two 100-nm-thick Ti coated glasses. Impedance measurements were carried out between 100 kHz to 0.1 Hz with an AC amplitude of 10 mV. The real impedance at the highest frequency was taken as the bulk resistance.

*Dopamine Detection Measurements*: Electrical measurements for device response were performed with a *Keithley 2612* and a Labview program in a 1x PBS, pH 7.4, buffer solution. Before sensing experiments, the device operation was confirmed and stabilized with transfer sweeps. The response of the sensors to the analyte was measured in constant bias mode with $V_D$ = -0.1 V and $V_G$ = 0.5 V and observing the drain current $I_D$. Transfer characteristics were obtained by applying $V_{DS}$ = -0.1 V and sweeping $V_{GS}$ from (0 to 0.55) V with a scan rate of 25 mV s$^{-1}$ until stabilization is obtained. Only the last complete cycle was examined to compare the sample before and after dopamine sensing. The buffer solution was changed multiple times to get a stable signal and a better basis for fitting the device drift before dopamine sensing. These multiple changes in the blank buffer solution can be fitted with an exponential decay fit.

*Photoelectron Spectroscopy Measurements:* XPS and UPS were performed in a UHV surface analysis system equipped with a Scienta-200 hemispherical analyzer. The excitation source for UPS was a standard He-discharge lamp with hν = 21.22 eV (He I), and for XPS, monochromatized Al Kα radiation with 1486.6 eV energy was used.

**Acknowledgments**
S.Z. thanks to Jianfei Li for writing the program to extract the cycle stability data reported in Fig. 2e. M.M thanks Chi-Yuan Yang for taking pictures of the OECTs on the orchid and bell pepper. This work was financially supported by the Knut and Alice Wallenberg Foundation, the Swedish Research Council (2020-03243), ÅForsk (18-313 and 19-310), the European Commission through the MSCA-IF-2020 project BEACON (GA-101024191), the FET-OPEN project MITICS (GA-964677), and the Swedish Government Strategic Research Area in Materials Science on Functional Materials at Linköping University (Faculty Grant SFO-Mat-LiU 2009-00971).

**Authors Contributions**
M.M. and S.Z. contributed equally. M.M designed, fabricated, and characterized the OECTs, and optimized the PEDOT:PSS and electrolyte formulations. S.Z. synthesized the GO, the cellulose/PVA formulation, and helped characterize the inks. Both M.M. and S.Z. contributed to the layout and writing of the paper. H.P. characterized the neuromorphic properties of the OECTs, while B.B. and C.D. performed the dopamine detection experiments, supervised by D.S. D.T. helped with the device layout and optimization. X.L. and M. F. performed the XPS/UPS characterization of the rGO. S.F. supervised the work and wrote the manuscript.

# Supporting Information

## Fully 3D-Printed Organic Electrochemical Transistors

Matteo Massetti[1], Silan Zhang[1,2], Harikesh Padinare[1], Bernhard Burtscher[1], Chiara Diacci[1], Daniel T. Simon[1], Xianjie Liu[1], Mats Fahlman[1,2], Deyu Tu[1], Magnus Berggren[1,2], Simone Fabiano[1,2]*

[1]Laboratory of Organic Electronics, Department of Science and Technology, Linköping University, SE-601 74 Norrköping, Sweden. E-mail: simone.fabiano@liu.se

[2]Wallenberg Wood Science Center, Linköping University, SE-601 74 Norrköping, Sweden.

Keywords: 3D printing, ink formulation, OECTs, organic mixed ion-electron conductors

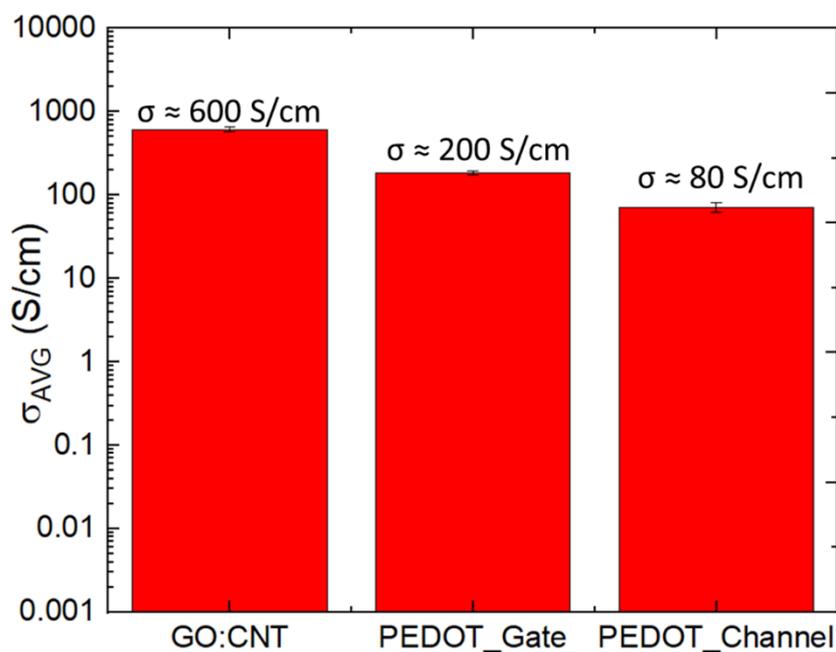

**Figure S1**. Electrical conductivity measurements, using the 4-point-probe collinear method, of the 2 PEDOT:PSS formulations and of the rGO/CNT electrodes. The electrical conductivity is calculated by $\sigma = \frac{l}{Rwt}$, where $l$ is the thickness, $w$ is the width and $t$ is the is thickness. $R$ is the extrapolated resistance.

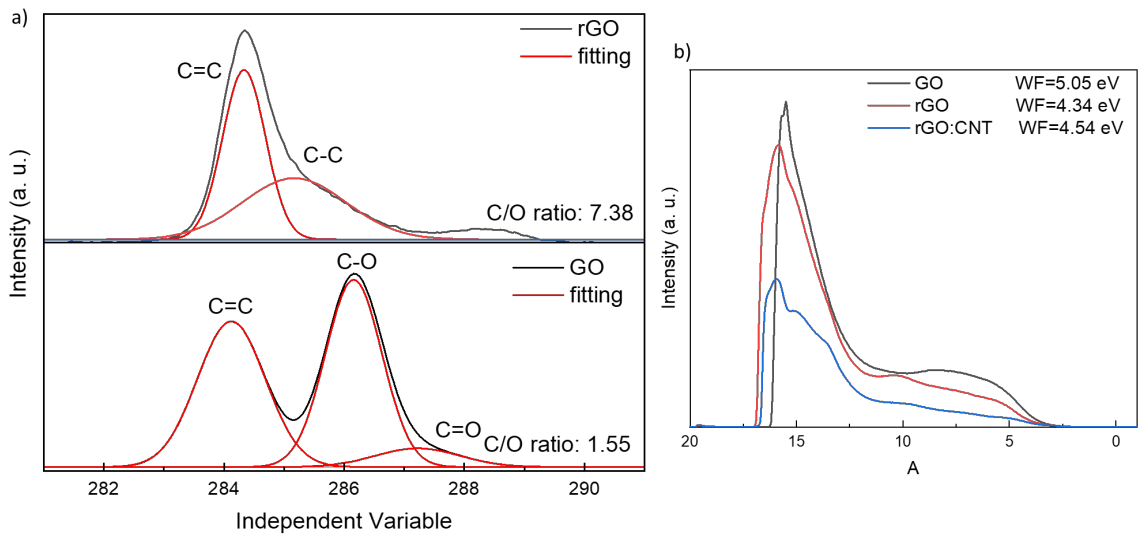

**Figure S2**. a) XPS measurements to determine the oxidation level of the GO electrodes before and after reduction treatment. It is possible to observe that the peak convolution does not show anymore C=O peak after reduction and the C/O increased from 1.55 to 7.38. b) UPS measurements to evaluate the WF of GO, rGO and rGO/CNT, showing the homogeneous dispersion of rGO with CNT.

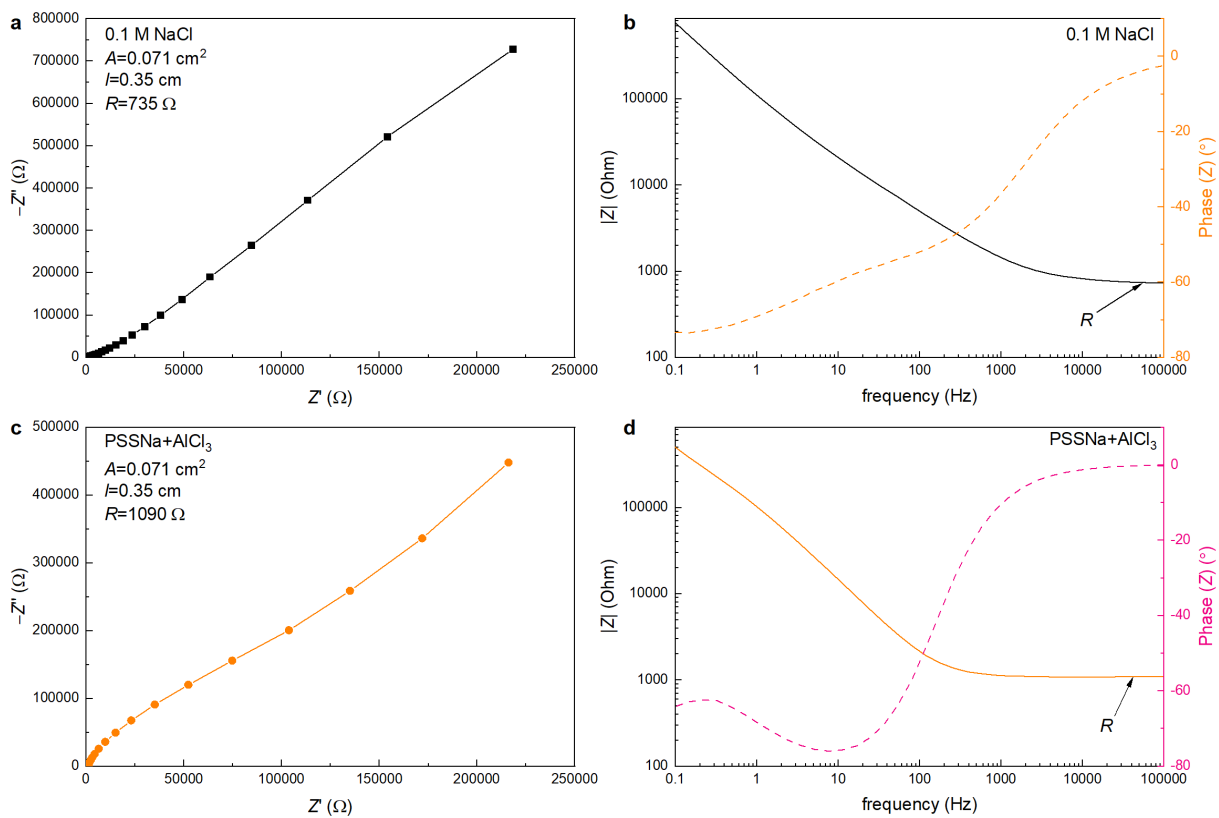

**Figure S3**. Ionic conductivity measurements for (a-b) 0.1 M NaCl and (c-d) PSSNa gel electrolyte. The ionic conductivity is calculated by $\sigma = \frac{l}{RA}$, where $l$ is the thickness and $A$ is the hole area of PDMS, $R$ is the bulk resistance, respectively.

Table S1. Performance comparison between this work and other PEDOT:PSS-based OECT

| | Techniques | $W/L/d$ (μm) | Gate Materials | Contact electrode | Substrate | electrolyte | $g_{max}$ (mS) | $I_{ON}/I_{OFF}$ | $\tau_{ON}/\tau_{OFF}$ (ms) | References |
|---|---|---|---|---|---|---|---|---|---|---|
| Planar techniques | Photolithography | 10/5/0.4 | Ag/AgCl wire | Au | Parylene on silica | NaCl (aq) | 2.7 | ~10 | 0.1/- | [28] |
| | Screen printing | 150/50/0.3 | PEDOT:PSS | Ag with carbon | PET | AFI VV009[a] | 1.6 | ~$10^5$ | ~30/~20 | [17] |
| | Screen/aerosol jet printing | 14.6/70.3/0.31 | PEDOT:PSS | Ag with carbon | PET | E003 | 1.1 | ~$10^4$ | 16/4 | [22] |
| | Ink-jet printing | 1000/3000/1 | Graphene/Ag | Ag | Polyimide | PBS (aq) | 1.2 | - | -/- | [19] |
| | Aerosol-jet printing | 200/200/0.2 | PEDOT:PSS | Ag | Kapton foils | NaCl (aq) | 0.5 | 1~10 | -/- | [23] |
| 3D-printing | Stereolithography/ 3D printing | 700/1900/500 | PEGDA:PEDOT | PEGDA:PEDOT | Spot-HT resin | NaCl (aq) | 2.5 | ~$10^3$ | 1130 | [33] |
| | Dip-coating on 3D printed structure | 12,000/15,000/- | PEDOT:PSS | Au | Nylon structure | PBS (aq) | 1 | 2 | 200000 | [34] |
| | Direct writing 3D printing | 1,000/1,600/7.1 | External Ag/AgCl pellet | Ag | PET or PLA | NaCl (aq) | 31.8 | ~$10^3$ | - | [6] |
| | Direct writing 3D/inkjet printing | 10000/3000/60 | External Ag/AgCl pellet | c-PLA[b] | TPC[c] | NaCl (aq) | 0.23 | ~2.5 | >2000/1000 | [24] |
| | Fully 3D printing | 240/77/0.5 | PEDOT:PSS | rGO/CNT | cellulose | PSSNa gel | 3.9 | ~$10^3$ | 100/27 | **This work** |
| | Fully 3D printing | 240/77/0.5 | PEDOT:PSS | rGO/CNT | cellulose | NaCl (aq) | 3.3 | $10^4$ | 7.8/7.5 | **This work** |

a. The electrolyte *AFI VV009*, which is based on poly(diallyldimethylammonium chloride) dissolved in water, solid particles providing opacity, initiator molecules and binder molecules capable of forming a cross-linked network through radical polymerization upon UV light irradiation, is provided by RISE Acreo on commercial terms
b. c-PLA: carbon-filled polylactide resin composite

c. TPC: thermoplastic co-polyester

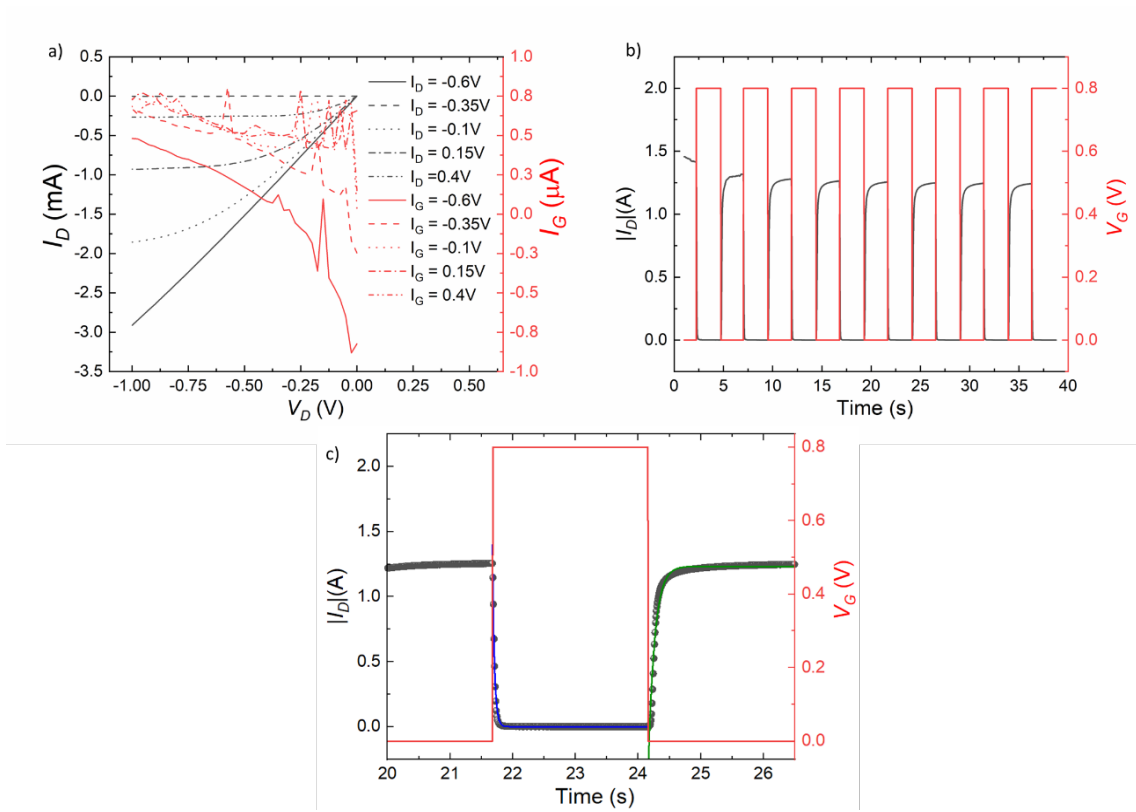

**Figure S4**. a) Output curve ($I_D$ vs $V_D$ @ different $V_G$) for the 3D printed OECT on parylene. b) $I_D$ response to $V_G$ pulsing with c) showing the exponential fitting to determine the time response.

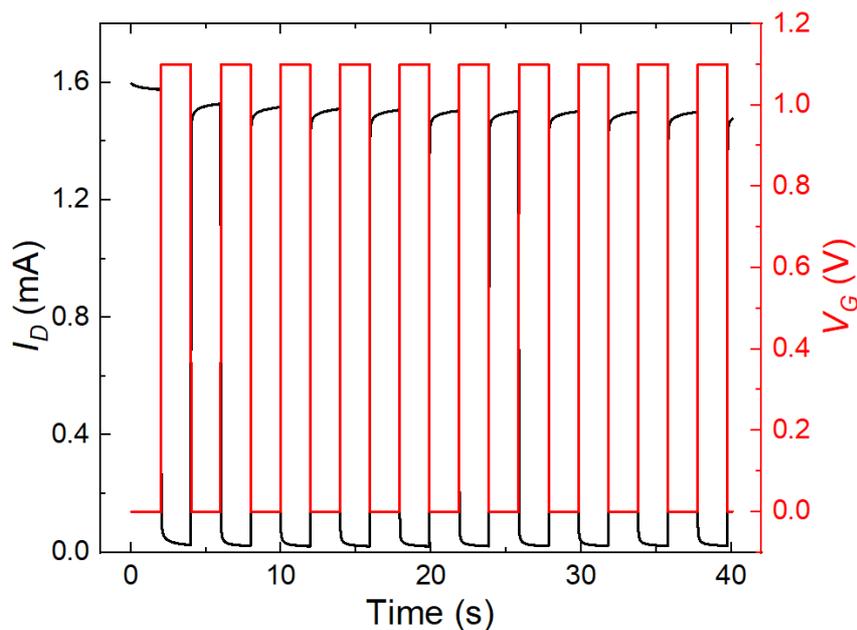

**Figure S5**. $I_D$ response to $V_G$ pulsing using 0.1M NaCl as electrolyte. The extrapolated time response $\tau_{OFF}$ and $\tau_{ON}$ are 7.5 and 7.8 ms respectively.

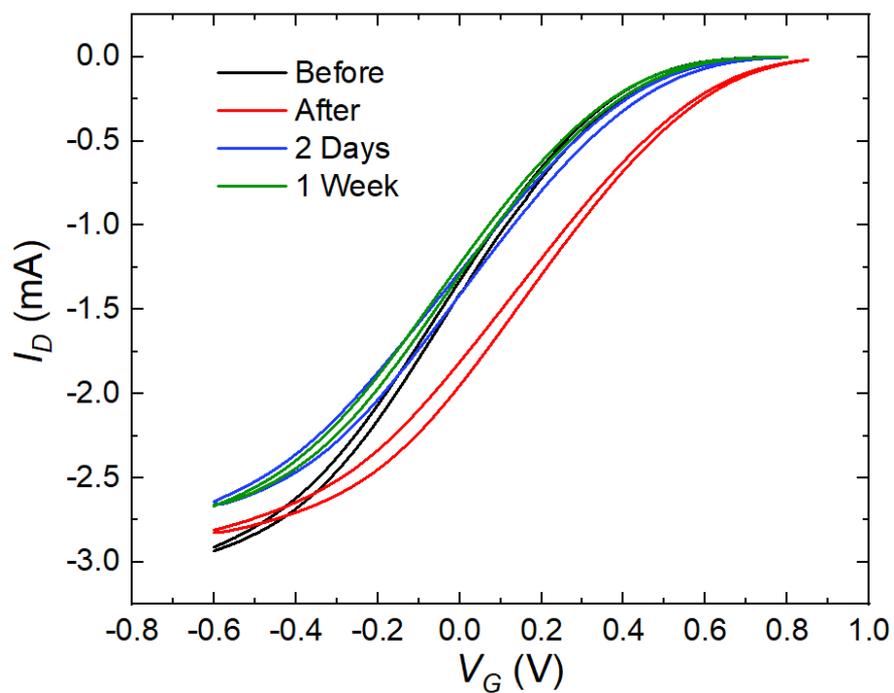

**Figure S6.** Comparison of the transfer curves before and after the cycling stability and their recovery after resting at 6 °C and high humidity.

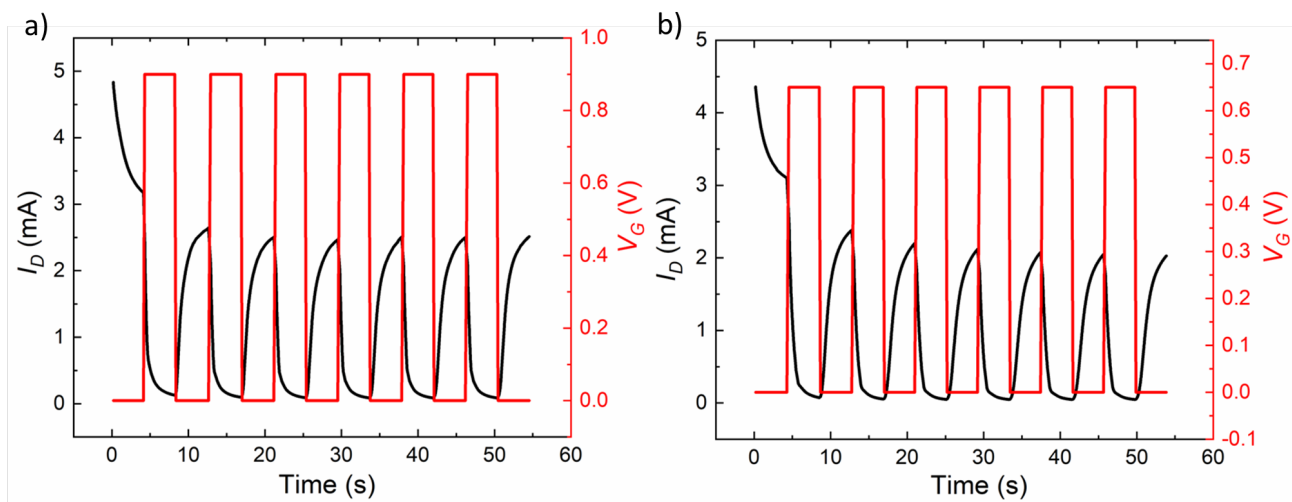

**Figure S6.** $I_D$ response to $V_G$ pulsing using the PSSNa gel electrolyte for OECTs printed on a) bell pepper and b) orchid.